Extended Inclusive Fitness Theory bridges Economics and Biology through a common understanding of Social Synergy

Klaus Jaffe

Universidad Simón Bolívar

Caracas, Venezuela

kjaffe@usb.ve, Tel +584129063610

**Abstract**

Inclusive Fitness Theory (IFT) was proposed half a century ago by W.D. Hamilton to explain the emergence and maintenance of cooperation between individuals that allows the existence of society. Contemporary evolutionary ecology identified several factors that increase inclusive fitness, in addition to kin-selection, such as assortation or homophily, and social synergies triggered by cooperation. Here we propose an Extend Inclusive Fitness Theory (EIFT) that includes in the fitness calculation all direct and indirect benefits an agent obtains by its own actions, and through interactions with kin and with genetically unrelated individuals. This formulation focuses on the sustainable cost/benefit threshold ratio of cooperation and on the probability of agents sharing mutually compatible memes or genes. This broader description of the nature of social dynamics allows to compare the evolution of cooperation among kin and non-kin, intra- and inter-specific cooperation, co-evolution, the emergence of symbioses, of social synergies, and the emergence of division of labor. EIFT promotes interdisciplinary cross fertilization of ideas by allowing to describe the role for division of labor in the emergence of social synergies, providing an integrated framework for the study of both, biological evolution of social behavior and economic market dynamics. For example, a utility function build on EIFT helps understand the appearance of terrorism and might help in eventually combating it.

# Introduction

The present paper does not pretend to present novel facts nor brand new theory. It aims at opening novel windows that allow for fresh views on established knowledge, favoring the flux of ideas between areas of science that have developed quite independently from each other. New multidisciplinary ways to look at old facts broaden our understanding of nature by helping us rethink established dogma in search of Consilience (Wilson 1999). Here I present a summary of a life long effort in building such a interdisciplinary window.

The theory of evolution, formulated by Darwin and Wallace some time ago, was built on the insight that heredity, natural selection, and variability interacted to produce biological evolution. The breakthrough in thinking was not the discovery of natural selection, or of heredity, or diversity. All these features were described in detail by Alexander von Humboldt, much cited by Darwin, and who lived a generation before Darwin and Wallace (Humboldt died the year Darwin published the *Origin of Species*). Humboldt had a working knowledge of selection and of the importance of the survival of the strongest, of heredity and the logic of domestication of plants and animals by selective breeding, and was aware about diversity, describing detailed variations between species and among species (Humboldt 1807). The important contribution by Wallace and Darwin was the insight that evolution emerged from the synergistic interaction of these three features, and that this evolutionary dynamics could explain the emergence of species (Top right cycle in Figure 1). That is, the continuous interaction between heredity, variations produced by mutations and the environment, and natural selection, produce the evolutionary dynamics that allows species to adapt to their environments and eventually diverge in their evolutionary path producing new species. That is, natural selection operates through the differential reproduction of individuals, measured as fitness. Higher levels of fitness are achieved by higher rates of reproduction, which in turn may be enhanced by higher survival probabilities. This theory, however, had difficulties in explaining many exceptionally bright colors, ornate plumage and conspicuous forms among living creatures, which attract predators and thus decreased the odds of individual survival. To overcome this limitation, Darwin introduced the concept of sexual selection to complement that of individual selection to explain biological evolution (Bottom right cycle in Figure 1). Darwin open two conceptually different ways at looking at fitness: individual survival that favors the strongest and most able individual, and sexual selection that favors the most prolific in mates and descendants. Both selecting forces might work synchronously, or they might diverge. Although both, survival and reproduction, are parts of the individual's fitness, theory that looks separately at each of the two processes help us in gaining a deeper understanding of evolution.

Soon after Darwin, important advances in our understanding about how evolution proceeds emerged. Development of population genetics, mainly between 1918 and 1932, and the expansion of Mendelian genetics were incorporated together with a more detailed theory of natural selection and gradual evolution into a modern evolutionary synthesis. This modern synthesis, produced between 1936 and 1947, reflects the consensus that is still valid today (Haldane 1932, Huxley 1942, Fisher 1958, Dobzhansky 1970, Mayr 1963, and others). The next important advance was a better understanding of cooperation and the emergence of societies that was not explained satisfactorily by Darwin (1859) and Wallace (1870) nor the just mentioned synthesis (Wilson 2000). Cooperation is important in a number of settings, including, behavioral interactions, biological evolution, sociobiology, cultural dynamics, and collective intelligence; yet the features allowing it succeed are not well known and are still discussed today (Skyrms et al. 2014). Inclusive Fitness Theory (IFT) as originally stated by Hamilton (1963, 1964), has been the most successful theory so far to provide explanations for the evolution of cooperation. Hamilton grasped that the effect of other individuals (con-specifics or not) affect the odds of survival of an individual. Specially among social species, the action of others could affect the fitness of an individual, so as to form a web of relations that affect the fitness of the participating organisms (The top left cycle in Figure 1 represents just one cell of such a network).

**Figure 1**: Schematic representation of selected aspects or components of the network of relationships responsible for the dynamics of natural **Selection** driving biological evolution. **Individual Selection (i)** represents natural selection acting on the individual; **Sexual Selection (s)** that acts on mate selection strategies and intra-sex competition; and **Inclusive Fitness (o)** cycles represents the coevolutionary effect on selection of the action of other organisms. **Variation** represents genetic mutations and phenotypic variations, **Reproduction** represents the reproductive and life-history strategies of individuals, **Mating** stands for sexual reproduction. Organisms suffer evolution through Individual Selection (bold arrows), which in turn is affected by at least two other cycles: Sexual Selection and Inclusive Fitness. Evolution among asexuals differs from this description (Jaffe 1996), as no mating's occur.

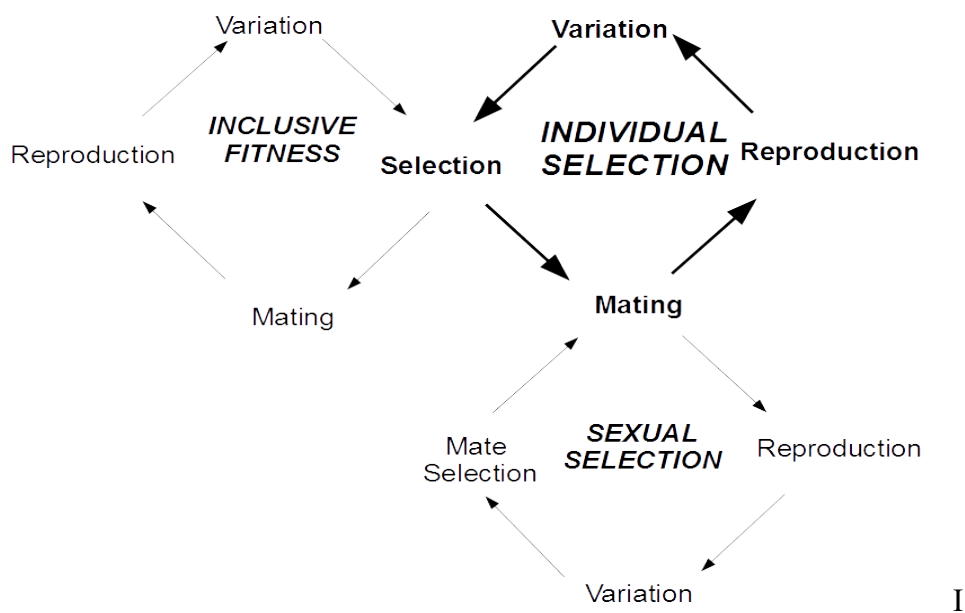

IFT can be summarized by an expression quantifying the fitness costs of a cooperative interaction as $c < b.r$; where "c" is the fitness cost to the donor in a cooperative act, "b" the benefit to the receiver, and "r" as the probability that an allele in one individual will also be present in a second individual via common descent. This simple formula is often misunderstood. The expression "c < b" is a consequence of the law for the conservation of energy, or first law of thermodynamics, as applied to biology: in the long term, survival of organisms requires that its total expenditures must be equal or lower than its total income. In order for fitness to be positive, positive survival rates are required. Hamilton's proposition was to treat "b" as a quantity modulated by "r". This IFT was often misunderstood, in part to the fact that soon after rejecting Hamilton's original paper on IFT submitted to Nature (Segerstrale 2013), Maynard Smith introduced the Kin Selection Theory (KST) to explain the phenomena Hamilton described with IFT (Maynard-Smith 1964). This historical circumstance has obscured the relevance of IFT until today and favored that of KST which is much more intuitive and easier to understand. The difference between KST and IFT is that the former only considers the genetic relationship between cooperating individuals as relevant for calculating fitness, whereas the later accepts that other factors are also relevant. KST simplifies IFT by assuming that "r" in the formula "$c < b.r$", represents "only" the genetic relatedness between donor and receiver. This simplification, though, has become very popular. So much so that Google Scholar in April 2015 retrieved about 1.4 more papers using the term "kin selection" compared to "inclusive fitness". Many scholars today still do not distinguishable between both concepts. (see Gardner et al. 2011, Allen et al. 2013, Corning 2013, for example). This confusion between KST and IFT has led several descriptions of the history of IFT to assign Haldane a primary role in it (see for example Hölldobler and Wilson, 2009). Yet, relating Haldane's (1932) casual comments on how expanded parental care may be favored by selection, with a pioneering role in the development of IFT, is equivalent to calling Alexander von Humboldt the grandfather of Darwin's theory of evolution. I argue that the components affecting inclusive fitness that are nor related to kin are much more important in explaining evolutionary phenomena, including economics, that those considered by KST. The trouble with KST is that it leads people to believe that cooperation can only be achieved when considering actions between relatives, whereas IFT can also explain cooperative interactions between non-kin.

Substituting IFT with KST was never accepted by Hamilton (Segerstrale 2013), and even Maynard Smith (1983) recognized its distinctiveness. IFT is a much more general theoretical framework than KST, which is a special aspect of the former. The focus on inclusive fitness rather than on kin selection allows for a finer understanding of population genetic dynamics. Inclusive fitness being > 0 can be the right criterion for social behavior to be selected, also in models where kin selection is absent, and where assortment is brought about by something other than common descent. Inclusive fitness, in addition to the genetic relatedness between the actors in a cooperation, takes into account "the likelihood of sharing genes above random levels due to statistical effects in genetic population dynamics" (Price 1970). I.e. the effect of co-variance on selection, that also determines the degree of assortation that may occurs between organisms (Price 1971). IFT is an open theoretical framework, which might conceive as multipliers to "b" any means that increase the frequency of an allele in a population through social interactions, such as mutualism, synergistic cooperation and others (Queller 1985, 1992, 2011). In fact, it is not necessary to refer any more to kin-selection. Flecher & Doebeli (2006) wrote: "The most fundamental explanation for how altruism (defined by local interactions) increases in a population requires that there be assortment in the population such that the benefit from others falls sufficiently often to carriers (and at the same time nonaltruists are stuck interacting more with each other). Nonadditivity if present can play a similar role: when collective cooperation yields synergistic benefits (positive nonadditivity) altruistic behaviour can evolve even in the absence of positive assortment, and when there are diminishing returns for cooperation (negative nonadditivity) the evolution of altruism is hindered (Queller, 1985; Hauert et al., 2006)."

Independently of the theoretical development just described, robust tools for handling non linear emergent phenomena in mathematical biology, such as numerical simulations, reached the same conclusions, confirming a central role for social synergy in the evolution of cooperation, specially among non-kin groups.  That is, agent based computer simulations studying the evolutionary dynamics of inclusive fitness on haploids, diploids, haplo-diploids, asexual and sexual organisms showed that social cooperation without social synergy is unable to emerge and sustain itself in scenarios for biological evolution (Jaffe 2001) and in scenarios of economic markets (Jaffe 2002a). These simulation showed that both, biological evolution of social behavior and market dynamics, require social synergy for its working. Social synergy is defined here as the process by which emergent properties arise through social interactions. For example, cooperation in retrieving food by insects allows them to handle food that no single individual would be able to capture and retrieve alone, expanding opportunities to exploit novel niches to groups of cooperating foraging workers. Such type of cooperation seems to explain the evolution of social behavior among bees (Michener 1969) and wasps (Silva & Jaffe 2002).  Social synergy is not reduced to an abstract concept as it can be measured quantitatively and empirically in different settings (Jaffe 2010). Cooperation were all interacting individuals benefit are also called Mutualism (Axelrod & Hamilton 1981, Bronstein 1994, Hoeksema & Bruna 2000, for example) and can be viewed as a special kind of social synergy.

EIFT is based on Queller's  version of Hamilton's rule (Queller 1985), as presented by Flecher & Doebeli (2006),  who formulated "r",  the modulator of "b" as a ratio of covariances (cov) so that:          $r = \text{cov}(G_A, P_0) / \text{cov}(G_A, P_A)$ where $G_A$ measures the genotype or breeding value in each individual in the population (subscript A for actor), $P_A$ the phenotypic value of each actor (e.g. 0 for defection and 1 for cooperation), and $P_0$ is the average phenotype of others interacting with each individual actor (subscript O for others).

This formulation implies that the altruistic genotype represented by $G_A$ increases in frequency if those with the genotype on average get more benefit from the behavior of others than they pay in cost for their own behavior. This relationship uses measures of assortment (covariance) between those with this focal genotype and the helping behaviours of others, scaled by the value of these behaviors. Taking the covariance over the whole population ensures that if this inequality holds for the helping genotype, it cannot simultaneously hold for the alternative nonaltruistic genotype. Therefore, when Hamilton's rule is satisfied, carriers on average have higher direct fitness than the population average (For details see Flecher & Doebeli 2006).

Here I propose a slightly different formulation that facilitates its application to human economic dynamics. This expanded theory allows to bridge conceptual divides between biological and economic sciences. Very recently, Corning (2013) presented a similar bioconomic approach to cooperation giving a preponderant role to synergy in evolution. His approach differs somewhat to the one developed here, as it focuses on multi-level and group selection (Corning 2013) and to 'synergistic selection' in the context of the emergence of complexity (Corning & Szathmáry 2015). The present proposition differs in its conception of inclusive fitness but can be viewed as complementary to Corning's approach, and is not a substitute for it.

## Expanding Inclusive Fitness Theory

A dynamic narrative that includes both biological and cultural evolution requires a few semantic modifications in order to formulate an EIFT. The first adaptation is to refer to agents instead of organisms. In biology, mating is described as a cooperation between two agents to produce offspring; whereas in economics, cooperating agents are productive units which can be individuals or aggregates such as companies. Cooperation is at the heart of any business and thus the basis of economic dynamics. Using agents as the unit for dynamic studies is getting more common in biology, sociology, ecology and economics as shown by the ever increasing literature (some examples are: Axelrod 1997, Pepper & Smuts 2000, Bonabeau 2002, Epstein 2006, Tesfatsion & Judd 2006, in addition to the literature cited so far).

Another semantic modification refers to reproduction. Reproduction should be viewed as reproduction of information, which includes diffusion and multiplication of information. This information can be of the genetic kind in biology, or , in economy it might mean memes (Dawkins 1989), information attached to productive systems (Hausmann & Hidalgo 2014), or scientific knowledge quantifiable with scientific papers (Jaffe et al. 2013b), etc.

As represented in Figure 1, the fitness of an individual has at least three aspects or components:
1- The selection acting on the survival capabilities of the individual that relate to its capacity to manage and respond to its environment (i),
2- The abilities to mate and reproduce that can be grouped under sexual selection (s),
3- The inclusive fitness or fitness affected by the presence and actions of other individuals with which it interacts or which it bestows upon others (o).

The total fitness of an individual (f) is a composite function which includes the fitness conferred by the phenotype of the individual, which in turn depends on its individual survival capabilities (i) and its reproductive success (s). In addition, f depends on the consequences of interactions with others (o), so that: $f = f(i, s, o)$

The component "o" has at least 3 parts to it. 1- The likelihood that a gene is present in another individual due to genetic relatedness or the kin selection component ($k$), 2- The probability that a gene is shared due to assortation ($a$), 3- The probability that a gene will favor the fitness of another due social synergies or economic considerations that emerge from the presence of specific alleles in each individual ($e$). Therefore: $o = f(k, a, e)$.

*A*ssortation includes the concepts of kin selection, as preference for cooperating with kin is a specific kind of assortation. Thus $o = f(a, e)$. The fitness of the individual f can be summarized as the product of two related networks of relationship : $f_i$ or factors affecting individual fitness directly; and $f_o$ or factors affecting the individual fitness via the action of others:

$$f = f_i(i, s, f_o(a, e))$$

This formulation converges with that supported by Queller (1985, 2011), Flecher & Doebeli (2006) and others, in that is treats assortation (*a*) and social synergy or non additive benefits (*e*) as the most important features determining the evolutionary viability of cooperation. Of all these terms, *a* and *e* are the less well understood and will explained below.

*Assortation:*

Assortation refers to the fact that similar organism attract each other. This is described in phrases such as "birds of one feather flock together" and is also refereed to as homophily: love for things similar to oneself; or narcissism: love of oneself. The term assortation was already used by Hamilton (1975) and he helped to show its relevance to IFT motivating George R. Price to develop a mathematically useful formulation (Price 1971). This paper showed that assortative mating can increase the frequency of an allele. This effect was shown to be so fundamental that it also works in mate choice in sexual reproduction (Jaffe 2002b). Complementing these findings, computer simulations showed that without some kind of assortative mating, sexual reproduction is unlikely to emerge among complex diplod organisms. (Jaffe 2001).

The working of assortation in favoring the success of cooperative strategies seem to be associated with the possibility of forming globular clusters, as is the case of some network structures (Kuperman & Risau-Gusman, 2012). Assortation is favored by tags or a green beard effect (Hamilton 1964), consisting of signals, behaviors or other features that allow agents to discriminate among potential cooperators and regulate the type of agents that will interact cooperatively (Riolo et al, 2001; Kim, 2010). A very basic form of cooperation often occurs among sexually reproducing mates. But assortation or homopyly evolves in many other cooperative interactions (Fu et al. 2012). Many behaviors of modern humans, such as the choice of mates and pets, can be explained as a result of assortation. For example humans select mates based on visual perception of their faces (Alvarez & Jaffe 2004), or of their pets (Payne & Jaffe 2005), and friends (Christakis & Fowler 2014) assortatively. In addition, homophily is very common in social settings (Centola et al. 2007, Kossinets & Watts 2009). Assortation or homophily have particular interesting effects on the evolutionary dynamics of cooperation, even beyond what can be explained with IFT. They reduce error thresholds of mutations (Ochoa & Jaffe 2006), and accelerate the speed of evolution (Jaffe 2001), favoring the emergence and maintenance of cooperation (Jaffe 2002b).

Assortation has been studied extensively in assortative mating and assortative cooperation (see review in Jaffe 2002b). Empirical evidence for assortation has been mounting. Here, just few random examples: Evidence among vertebrates include studies showing that chimpanzee friendships are based on homophily in personality (Massen & Koski 2014); the existence of assortative mating in lesser snow geese (Cooke et al. 1976) and in blue tits (Andersson & Andersson 1998); assortation among humans in games of experimental economics (Bowles et al 2005, Hamilton & Taborsky 2005); from anthropological and archaeological studies (Apicella et al. 2012, Richard & Levin 2005); and of course, from ethology (Alvarez & Jaffe 2004, Payne & Jaffe 2005).

***Social Synergy:***

Much work on cooperation has centered on altruism. Indeed, the ultimate sacrifice of ants and bees for the good of their colonies is an impressive feat. But eventually, all sustainable social behaviors involve interactions that are beneficial to all intervening parts (Flecher & Doebeli 2006). Interactions where all parties gain, the so-called win-win interactions (Figure 2), are very much known among economist (Lewicki et al. 1985, Dolfsma & Soete 2006, Liu & Huang 2007). One important concept is Social synergy, i.e. non-additive benefits and positive feedback of social behavior that affect individual fitness. Social synergy refers to synergies triggered by social cooperation that increase economic and other benefits to social individuals favoring its evolution (Queller 1992, 2011, Jaffe 2001, 2002, 2010, Taylor 2013). Synergies that emerge from social interactions can be quantified (Jaffe 2010, Bettencourt 2013), and are fundamental in explaining the maintenance of complex societies (Jaffe 2002a).

Although not ignored, social synergy has been little studied quantitatively among living creatures other than humans (Jaffe 2010). The father of IFT already recognized that several different mechanisms are needed to explain the prevalence of social cooperation among extant species (Hamilton 1996). Studying bees, Michener (1969) demonstrated the existence of several different evolutionary routes leading to sophisticated societies that benefited all or most of its members. To understand these evolutionary dynamics, economic and ecological considerations are more important than genetic ones (Osborn & Jaffe 1997, Silva & Jaffe 2002).

The economic forces unleashed by human cooperation have been studied by political economist for many years (Krotopkin 1902 for example). More recently, Social synergy has been mentioned when studying pay off matrices, altruistic punishment, benefits of social life and cooperation The cost/benefit ratio of cooperation might reveal the existence of this synergy. Cost/benefit ratios have been shown to be important for the evolution of cooperation in different settings (Nowak & Sigmund, 1998; 2005; Jaffe, 2002: Nowak, 2006; Baranski et al, 2006; Ohtsuki et al, 2006; Jaffe & Zaballa, 2010, Taylor 2013).

**Figure 2**: Effect of different cooperative strategies on the fitness of the actors.

In interactions involving exploitation or parasitism, one organisms benefit at the expense of the other, increasing its fitness (bigger dark blue faces) and reducing that of the other (small light blue face). In altruistic interaction the reverse hold. The altruist reduces its fitness and the other increases it. Interactions where altruists punish individuals not complying with social norms, both the altruist and the other reduce their fitness. In synergistic business both actors win, increasing their fitness, not necessarily by the same amount (Jaffe 2014a).

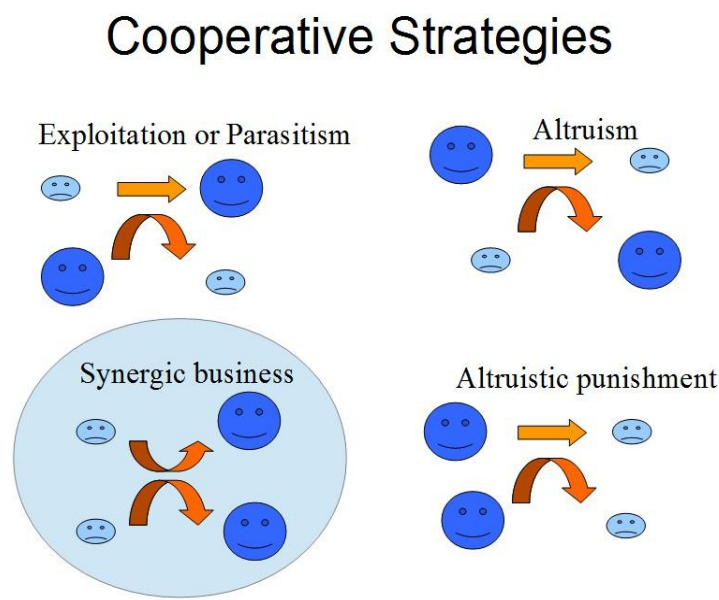

From where does this synergy that produces win-win situation arises? Economics has an answer to this question. Long ago, among others, Aristotle recognized that division of labor enlarges and elicits innate human differences (Aristotle IV BC), which allow the existence of complex society. Adam Smith recognized the existence of a special synergy working in the markets as an "invisible hand", but neither he nor others focused on the specific mechanisms that allowed its working. Adam Smith (1776) writes in The Wealth of Nations "The greatest improvement in the productive powers of labor, and the greater part of the skill, dexterity, and judgment with which it is anywhere directed or applied, seem to have been the effects of the division of labor .... It is the great multiplication of the productions of all the different arts, in consequence of the division of labor, which occasions, in a well governed society, that universal opulence which extends itself to the lowest ranks of the people". Friedrich Hayek (1949), specially when tackling the "Economic Calculus", and many others (Becker & Murphy 1994 for example), hint to the emergence of synergistic effect in social interactions due to the existence of differently specialized actors. Division of labor has also been associated with social life in insects and many other animals (Wilson 2000) and is correlated with the degree of order in ant societies (Jaffe & Hebling-Beraldo 1993, Jaffe & Fonck 1994). Thus, a very important source of social synergies in the economic dynamics of markets and in biological cooperation is the division of labor. The force behind Adam-Smith's invisible hand of the market that triggers the complex market dynamics intuitively described by Hayek, is social synergy as described here. This can be evidenced using computer simulations (Jaffe 2015) and robot swarms (Ferrante et al. 2015). It is the specialization of labor that allows complementary interactions to produce ever stronger synergies that confer non-linear economic advantages to societies that allow and foment individual liberty and division of labor (Jaffe 2014a).

Not all social synergies arise from division of labor. Many other mechanisms are possible. Economies of scale, for example, also allow the individual to achieve higher fitness or economic gains among humans and other animals (Hamilton 1971, Jaffe 2010, Bettencourt 2013). Social intelligence can also be viewed as an emergent phenomena (Woolley et al. 2010). Again, economic science has explored these issues much more and/or differently than biology.

Among economists, the existence of non linear dynamics in wealth accumulation has been recognized long ago. Karl Marx (1867) for example, when he described the surplus value and attempted to balance wealth across a society, recognized that more than simple additive arithmetic's was required. More sophisticated thermodynamic approaches to study non-linear dynamics in economics were initiated by Georgescu (1970) and developed further by many others (see Beinhocker 2006 for example) which eventually lead to a systematic use of the concept synergy in economics. Underlying the concept pf social synergy in economics is the fact that some actions and the exploitation of some resources is only possible after a certain threshold size of social aggregates has been reached, producing a non-linear or emergent effect on wealth aggregation.

Examples of synergy used to simulate social evolution in biology can easily be applied to human economics. The example of two wasp mothers that attend their brood communally, each one investing 50% of their time in brood care, achieving to protect their brood 100% of the time, reducing the odds of losing their brood to zero with the same cost to parents (Jaffe 2001), can be expanded to human societies and institutions in charge of communal security (Zabala & Jaffe 2010).

A simple example of a synergistic view of the relationship between increased utility and increased wealth is that if a wealthy donor gives a poor recipient a blanket, the recipient will get a much higher utility from the blanket than the donor, but there is no net increase in wealth. But if the object donated is a sewing machine, which is used in the rich donor's house as decoration, but the poor receiver uses it to produce blankets to sell, then there is a net increase in wealth. The first case illustrates a synergistic increase in utility, the second type one of wealth. (Libb Thims personal communication).

The more we look at synergy in economics and business management, there more we find meaningful examples. Examples include: the impact of acquisitions on merging and rival firms (Dopfer 1991, Chatterjee 1986), economic development (Evans 1996, Ostrom 1996), mergers and acquisitions (Bradley et al 1983, Fun et al 1996, Larsson & Finelstein 1999), and evidence that certain type of competition over personal resources can favor contribution to shared resources in human groups (Barker et al 2013)

## From Biological Evolution to Economic Dynamics

How relevant is EIFT for our understanding of evolution? Cooperation among non-kin is as or more important that between kin. For example, symbioses are far more important in biological evolution than hitherto recognized (Kiers & West 2015, Corning & Szathmáry 2015). Theoretical evolutionary theory needs to digest this fact. In addition, recent reviews provide ample theoretical and empirical evidence justifying the extensions to IFT. For example, Van Cleve and Akcay (2014) showed that the interaction between behavioral responses (reciprocity), genetic relatedness, and synergy interact are fundamental in understanding the richness of social behavior across taxa. The review by Bourke (2014) on "comparative phylogenetic analyses show that cooperative breeding and eusociality are promoted by (i) high relatedness and monogamy and, potentially, by (ii) life-history factors facilitating family structure and high benefits of helping and (iii) ecological factors generating low costs of social behavior". The last factor is of course the mirror image of social synergy: Environments provide selection pressure to which organisms evolving cooperative strategies producing social synergy has to adapt. Many unequivocal examples of social synergy as a factor in determining the evolutionary success of social behavior have been reported. The best known example is probably the evolutionary history of social behavior among bees (Michener 1969). In the case in wasps, Hamilton's preferred species, social behavior generates indirect benefits by enhancing the productivity or survivorship of non-kin more often than that of kin (Strassmann et al. 1991, Itô 1993, Gadagkar & Gadagkar 2009, Jaffe & Silva 2002, for example).

Biologists are not the only ones interested in social evolution. The features that influence the dynamics of cooperation have been studied using different theoretical frameworks with different specific assumptions. The theoretical framework of studies of social dynamics by biologists, sociologists, economists, physicists, mathematicians, game theorists, computer scientists, and others, differ in the concepts they use despite the fact that all are studying the same basic phenomena, making interdisciplinary communication of this issues difficult. However, all these disciplines have used applications of game theory, and specifically the Prisoner's Dilemma, to pursue their quest for answers in their fields. Thus, a common language bridging the concepts between these disciplines seems possible.

An important difference between biology and economy is that biology focuses on genetic evolution whereas economy studies cultural processes. This difference is much less important tan the homologies in dynamic processes. For example, Manfred Eigen (1971) insists that Darwinian evolution is not merely the organizing principle of biology but a law of physics that should be responsible for many phenomena in nature. Genetic evolution is based on vertical transmission of information, from parents to offspring, whereas cultural transmission includes in addition to the vertical kind a horizontal transmission of information. The overall evolutionary dynamics of both processes, however indistinguishably (Jaffe & Cipriani 2007). Both processes produce a continuous dynamics that may induce divergence or specialization (Jaffe et al. 2014).

The evolutionary dynamics in biology is centered on genes and organisms, whereas the economic dynamics is centered on business, enterprises and companies. In biology, mating or cooperation between the sexes is fundamental for the survival of the population; whereas in economy it is cooperation among different type of laborers or companies that allows production of wealth. In both cases, the dynamics driving information, innovation and social synergy is similar. EIFT formulas the equation: **c < r . b** using $r = f(a,e)$ or a function of the probability of the individual to posses a gene that confers it advantages in social interactions with others and the social synergy triggered by this interaction. In the case where the socially advantageous gene is shared between interacting organisms, we speak of assortation. If $f(a,e)$ has evolved by natural selection, economic utility functions are expected to include these same factors. Any comprehensive utility function in economics then has to include besides direct benefit/cost considerations, benefits to kin and group members, to individuals which might reciprocate positively in the future, and any entity that might synergize the individual actions. That is the utility function (u), analogously to the fitness function defined above, has to have at least three different components: $u = f(i, a, e)$.

*Benefits of an Extended Inclusive Fitness Theory?*

The challenge of EIFT is to explain in more detail how biological and economic systems produce synergies by favoring specialization and division of labor, conferring the individuals in a cooperative society with fitness benefits that are much higher compared to a solitary life (see Jaffe 2015 for example). More experimental approaches in economics are required to address these issues (see Tollefson 2015 for example) .

EIFT considers that factors other than genetic relatedness affect the cost benefit balance of cooperation and that fitness functions and utility function have to consider the direct effects on the individual as well as indirect benefits an individual achieves through assortation and synergistic interactions. These factors have been studied with different emphasis by biologist and economist. The most important factor often overlooked so far is probably the social synergy that emerge from cooperative interactions, such as synchronized division of labor. An important conclusion from empirical studies in economics, that try to assess the effect of social synergy or economic benefits that derive from social life, is that synergy is probably the most important driver in the evolution of cooperation, and that assortation or genetic relatedness are neither necessary nor sufficient for the emergence of cooperative phenomena (see also Corning 2013 ). The same conclusion is reached when exploring the dynamics of cooperation in the repeated prisoners dilemma game (Montoreano & Jaffe 2013). Here social synergy is more important than assortation, which in turn is more important than kin selection, in fomenting cooperation. This suggests that an expanded version of IFT is required for a better understanding of the dynamics. Focusing only on kin selection is not enough. An insight into the economics of the cooperation is fundamental in understanding it. However, little quantitative empirical research on social synergy has been produced in biology (but see Osborne & Jaffe 1997, Jaffe 2010, Smith et al. 2010), though it is recognized as of primary importance in the economics and business literature. The latest reviews of the empirical literature in biology, confirm that a more economic view explains the descriptions of societies found in nature better (Van Cleve & Akcay 2014; Bourke 2014). Even on co-evolution, the review by Ivens (2014) shows a pattern among farming mutualism of ants and their domesticated species that seems to produce stability of these successful mutualisms: The component of inclusive fitness in the evolutionary dynamics (Figure 1) dwarfs the sexual selection component. Most of these mutalisms are characterized by reduced symbiont dispersal and diversity (often in association with asexual reproduction and vertical transmission), promoted by specific ant behaviors of the ants, such as creation of protective environments. Coevolution, viewed in this new light (see Dawkins & Krebs 1979, Jaffe & Osborn 2004, Zaman et al. 2014, for example), can easily explain many symbioses. Even extravagant proposals such as the one stating that host-microbe interactions influence brain evolution and development in mammals (Stilling et al. 2014), can now be explained. An EIFT makes it unnecessary to treat symbioses and social cooperation as different phenomena as done by Corning (2013), as both are considered in o = $f(a, e)$.

The central insight from recent empirical studies is that economic factors and assortation in its different forms determine social behavior. Social behavior cannot be understood without taking account of all of them. A synergistic interchange of theoretical knowledge between economics and biology looks promising for a novel attempted to deepen our understanding of social dynamics and should help to bridge the gaps in studies of evolution of social cooperation between economist, physicists, biologists, and others, providing for a common language in the quantitative assessment of the importance of specific features that aid social evolution.

A theory that helps us to look for the relevant features in the evolution of social behavior, dynamics of cooperation and evolution of society might be useful. That is, more important than kin relationships are assortation and social synergy for understating social cooperation. Assortation is important in a number of fundamental instances of human cooperation (Jaffe 2002b, 2008, Weisel & Shalvi 2015) and may emerge in many other circumstances if we look for it. The most relevant potential contribution of this theory is that it might allow social science to profit from both economics and biology. It might help develop complexity sciences aiming to improve our understanding of social synergies unleashed by cooperation are of the fundamental forces driving the evolution of societies. These phenomena should be empirically observable. Three examples might help convince the reader about the empirical usefulness of this theory.

1- Many butterflies have associations with ants. They can either be mutualistic, exploitative or parasitic. Quantitative phylogenetic analysis revealed a large prevalence of cooperation over competition in the symbiotic relationship (Osborn & Jaffe 1997). As no possibility of genetic flow between ants and butterflies exist, there is no doubt here that social synergy is the driving force for cooperation. An impressive large number of symbioses are known to exist (Corning 1983, 2013). This unified treatment of social synergy can be expanded to address the spontaneous commerce and cooperation networks that arise from the working of competitive advantages between nations (Porter 2011) and firms (Grant 1991) in economics, as originally conceived by David Ricardo (1891).
2- Empirical evidence shows that different forms of division of productive activities in an economy accounts for differences in its capacity to produce and accumulate wealth (Hausmann & Hidalgo 2014). This is linked to division of intellectual labor (Jaffe et al. 2010) in contemporary human society. For example, the division of labor in academic research accounts better for differences in relative economic success among nations than any other variable studied so far (Jaffe et al. 2013a,b). These examples show that arrangements that affect social synergies, such as division of labor, are the key to understand contemporary economic development, including the working of finance (Jaffe & Levy-Carciente 2004). An EIFT stimulates the exchange of analytical tools between economists and biologists for a novel view of the working of division of labor (Jaffe 2014b, for example).

3- From a biological point of view, division of labor in ants is related to increased economic gains of social behavior (Jaffe & Hebling-Beraldo 1993) and at the same time, more sophisticated social behavior is related to a decreased individual complexity (Jaffe & Perez 1989). This is an example of social synergy driving social evolution at the expense of individual selection, easily explainable with the EIFT. In economics, we accept that societies confer energetic benefits to all individuals involved in both ants and humans (Jaffe 2010, Bettencourt 2013). Thus cities allow synergies to emerge that provide non-linear benefits to society (Haken & Portugali 2003, Florida 2005). This synergies are practically everywhere (Corning 1983) and can even be detected in basic physical architectural arrangements (Fuller 1975) and therefore might be present in many as jet unsubscribed situations (Haken 1973). An EIFT might be better able to develop analytical tools to understand how and why synergies emerge from division of labor (Jaffe 2014a).

4- The general utility function proposed here, that aims to maximize individual benefits directly and indirectly through assortation and synergistic interactions, has molded instincts and behavior in all extant plants and animals, including humans, as it is the product of natural selection. It explains features of modern life that has escaped explanations by classical economic theory. Terrorism for example is a feature that is ever more important in contemporary society. Motivations to commit terrorist acts, however, are driven by biological and economic stimuli. The branch of biology studying animal and human behavior, ethology, tells us that aggression enhances group cohesion, that poor odds of survival or of alternative routes to increase ones fitness (or utility) function increase the likelihood of aggressive interactions, and that differences in individual strength or low odds of retaliation favor aggression (Lorenz 1963, Eibl-Eibesfeldt 1979). That is, the consequences of a behavioral action can be assessed by the ratio of benefits (b) to costs (c). If $b/c$ is high, biological and economic evolution will favor this behavior. As recognized by ethologists, sometimes b tends to infinity allowing for the existence of supernormal stimuli (Tinbergen & Perdeck 1950, Mirás et al. 2007). Religion favors such a hyper-stimulus as shown by simulations (Jaffe & Zabala 2009, 2010). When pursuing heavens, or avoiding hell, benefits tend to infinity. Thus any action which guarantees heaven or immortality, maximizing the individuals utility function, will be favored, even if it implies self-immolation. This motivations, together with technological means that allow a single individual to inflict harm to many people, and to pass on virtual or real benefits to a much larger number of individuals belonging to his/her natural group, allow terrorism to proper, regardless if it is based on religious beliefs or not.

These examples show how a unified view of the dynamics governing cooperation might help achieving a better understanding between biology, sociology, economy, complex system sciences among others, eventually unleashing synergies that might advance our understanding of nature in important ways. Low hanging fruits might be found by economist exploring the working of assortation, which might improve our scant understanding of the interactions between family and business (The Economist, 2015). Homophily in human society achieves less diverse but more harmonious economies (Wang & Steiner 2015), suggesting a role for assortation hitherto overlooked in economics. Assortation, viewed in the light of the present theory might, for example, explain the ubiquity of corruption among human societies, and help biologists to better understand economic synergies found in the social phenomena they study, opening our interdisciplinary world view in a consilient way.

**Acknowledgments**: Thanks are due to Michael Hrncir, Cristina Sainz, Emilio Herrera, Rodolfo Jaffe, and Peer 1038 of Peerage of Science. The paper was written as a response to questions and discussions triggered by my plenary talk honoring the 50 years of Bill Hamilton Inclusive Fitness

Theory at the International Ethology meeting organized by the Sociedade Brasileira de Etologia in Mossoro in 2014. A non-quantifiable number of people helped crystallize the ideas presented during the last few decades. These include long evenings with Bill Hamilton during the eighties in the Americas and later in his house in Oxford; intensive conversations with John Maynard Smith in the University of Sussex a few months before his death; conversations and correspondence with David Queller; suggestions from Peter Corning; among many others.

Alternative Figure 2

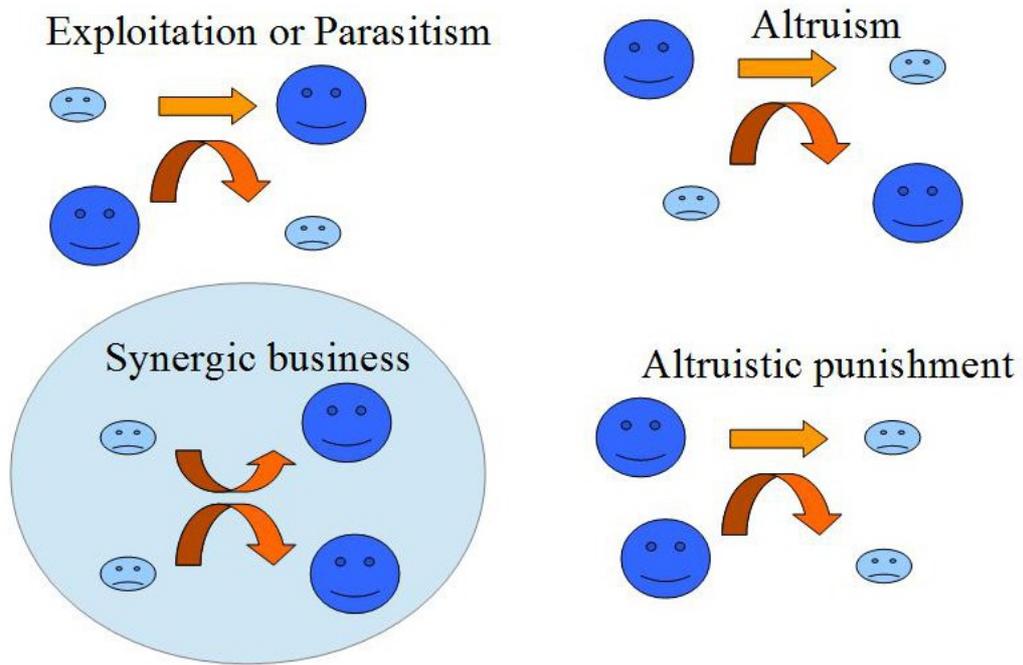